\documentclass[journal]{IEEEtran}
\usepackage{cite}
\usepackage{graphicx}
\usepackage{url}
\usepackage{amsmath}
\usepackage[caption=false, font=footnotesize]{subfig}
\usepackage{float}
\usepackage{booktabs} 
\usepackage{multirow}
\usepackage{array}
\usepackage{algorithm}
\usepackage{algpseudocode}
\usepackage{bbm}
\usepackage{dsfont}
\usepackage{tabu}
\newcolumntype{L}[1]{>{\raggedright\let\newline\\\arraybackslash\hspace{0pt}}m{#1}}
\newcolumntype{C}[1]{>{\centering\let\newline\\\arraybackslash\hspace{0pt}}m{#1}}
\newcolumntype{R}[1]{>{\raggedleft\let\newline\\\arraybackslash\hspace{0pt}}m{#1}}

\usepackage[table,xcdraw]{xcolor}
\usepackage{xspace}
\usepackage{mystyle}
\ifCLASSINFOpdf
\else
\fi
\begin{document}
%
\title{ER-IQA: Boosting Perceptual Quality Assessment Using External Reference Images}
%
%
%

\author{Jingyu~Guo,
        Wei~Wang,
        Wenming~Yang,
        Qingmin~Liao,
        and~Jie~Zhou


\thanks{J. Guo is with the Shenzhen Key Laboratory of Visual Image Processing, Shenzhen Internatinal Graduate School, Tsinghua University, Shenzhen 518055, China, also with the Department of Electronic Engineering, Tsinghua University, Beijing 100084, China (e-mail: gjy19@mails.tsinghua.edu.cn).}

\thanks{W. Wang was with the Shenzhen Key Laboratory of Visual Image Processing, Shenzhen Internatinal Graduate School, Tsinghua University, Shenzhen 518055, China, also with the Department of Electronic Engineering, Tsinghua University, Beijing 100084, China. He is now with the AI Lab, ByteDance, Beijing 100000, China (e-mail: wangwei.frank@bytedance.com). }

\thanks{W. Yang and Q. Liao are with the Shenzhen Key Laboratory of Visual Image Processing, Shenzhen Internatinal Graduate School, Tsinghua University, Shenzhen 518055, China (e-mail: yang.wenming@sz.tsinghua.edu.cn; liaoqm@tsinghua.edu.cn).}

\thanks{J. Zhou is with the Department of Automation, Tsinghua University, Beijing 100084, China (e-mail: jzhou@tsinghua.edu.cn)}

}

%
%

\markboth{}%
{Shell \MakeLowercase{\textit{et al.}}: Bare Demo of IEEEtran.cls for Journals}
%



\maketitle

\begin{abstract}
Recently, image quality assessment (IQA) has achieved remarkable progress with the success of deep learning. However, the strict pre-condition of full-reference (FR) methods has limited its application in real scenarios. And the no-reference (NR) scheme is also inconvenient due to its unsatisfying performance as a result of ignoring the essence of image quality. In this paper, we introduce a brand new scheme, namely external-reference image quality assessment (ER-IQA), by introducing external reference images to bridge the gap between FR and NR-IQA. As the first implementation and a new baseline of ER-IQA, we propose a new Unpaired-IQA network to process images in an content-unpaired manner. A Mutual Attention-based Feature Enhancement (MAFE) module is well-designed for the unpaired features in ER-IQA. The MAFE module allows the network to extract quality-discriminative features from distorted images and content variability-robust features from external reference ones. Extensive experiments demonstrate that the proposed model outperforms the state-of-the-art NR-IQA methods, verifying the effectiveness of ER-IQA and the possibility of narrowing the gap of the two existing categories.  


\end{abstract}


%
\IEEEpeerreviewmaketitle

\section{Introduction}
%
%
%
%
\IEEEPARstart{W}{ith} the explosively increasing number of digital images produced every day, assessing image quality subjectively has become more and more time-consuming and laborious in practical applications. Hence, objective image quality assessment (IQA) is in great need to automatically assess image quality when applied to image processing and computer vision tasks such as image generation \cite{ledig2017photo}, image restoration \cite{banham1997digital}, image retrieval \cite{guo2017robust, yan2014learning}, etc.

Current IQA approaches are generally divided into three categories, \ie, full-reference IQA (FR-IQA), reduced-reference IQA (RR-IQA), and no-reference IQA (NR-IQA) based on how much information of the undistorted image (also referred to as the reference image) is available during the quality assessing process. Though FR-IQA methods have achieved remarkable progress over the decades, these approaches are usually infeasible in practical applications when the pre-condition, \ie, requiring a corresponding reference image for comparison, is not satisfied \cite{lin2018hallucinated}. In contrast, NR-IQA is closer to the real scenario and has received substantial attention in recent years.

\begin{figure}[t]
\begin{center}
    \includegraphics[width=0.95\linewidth]{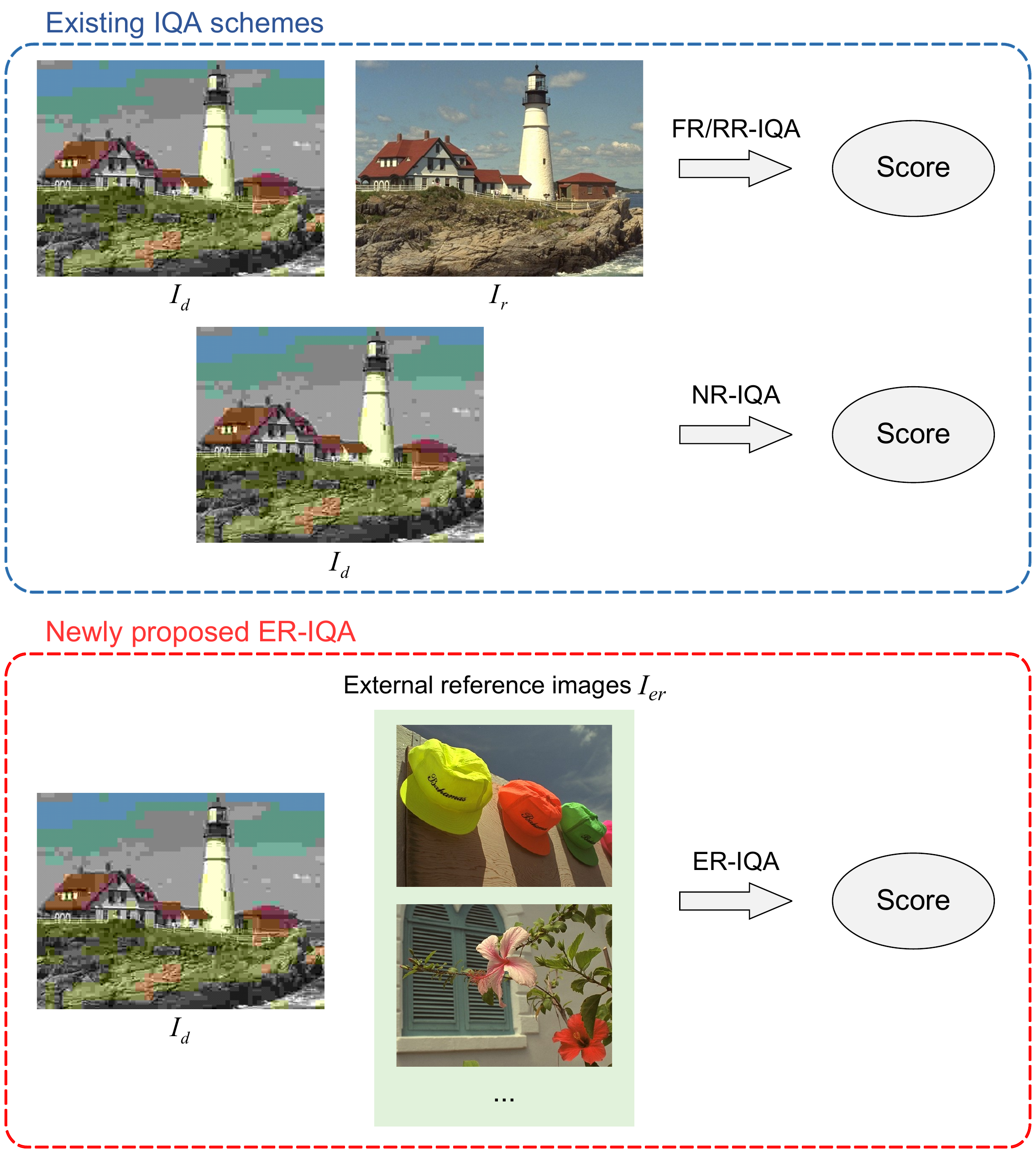}
\end{center}
   \caption{An illustration of our motivation. The existing NR-IQA scheme suffers from limited available information and an ambiguous definition of image quality, yet FR-IQA restricts the content of reference images to be identical to the distorted ones. Our approach exploits the potential of user-supplied external reference images and bridges the gap between the existing categories.}
\label{fig:introduction}
\end{figure}

Early NR-IQA approaches \cite{moorthy2010two}, \cite{moorthy2011blind}, \cite{saad2012blind}, \cite{mittal2012no}, \cite{zhang2015feature}, \cite{xue2014blind} tend to extract hand-crafted features from images and perceive quality scores by regression. However, these methods usually lack generalization ability since the designed features can hardly describe the multiple complex distortions in a wide range of images. Thanks to the powerful representation ability of convolutional neural networks (CNNs), CNN-based NR-IQA methods \cite{kang2014convolutional}, \cite{kang2015simultaneous} have achieved significant improvements compared to previous hand-crafted approaches. Most of these methods treat NR-IQA as a regression task and solve it in an end-to-end manner. Achieved promising results as they have, they suffer from the lack of large annotated datasets, a common problem in many deep learning tasks. Moreover, we argue that what makes the problem even more severe to NR-IQA is the insufficient supervision the training labels can provide. Unlike those in other vision tasks that have certain physical meanings, \eg, the category \cite{krizhevsky2012imagenet} or the position \cite{pascal-voc-2007}, \cite{redmon2016you} of objects, the labels in IQA datasets, \ie, mean opinion scores (MOSs) \cite{sheikh2006statistical}, \cite{ponomarenko2013color}, can provide very limited information. The definition of image quality is rather subjective, and even humans cannot tell the exact relation between the attributes of an image and its corresponding MOS, let alone algorithms. As a result, it is difficult for networks to learn quality-aware feature extraction under MOS' supervision alone. Existing approaches \cite{bianco2018use}, \cite{wu2015blind}, \cite{kim2016fully}, \cite{ma2017end} tend to tackle the problem through knowledge transfer, where different pre-training techniques are introduced and have been proved effective. One other solution is to manually extract prior information, mainly low-level features such as gradient maps \cite{yan2018two}, as an additional input. The common idea of these approaches is to enhance the physical meaning of image quality by introducing additional information. And once the models get enough information to understand image quality, they can extract quality-discriminative features and obtain high performances. 

We believe the huge gap between FR and NR-IQA is mostly due to how much information the algorithms can utilize when understanding and integrating image quality. In the full-reference scenario, the reference images serve as the upper bound for perfect quality images, making it possible for algorithms to simply compute the similarity between images and project it into quality scores. However, NR methods can utilize no such information, resulting in great performance degradation, which motivates us to narrow the gap by borrowing the idea of FR-IQA and setting the upper bound manually. As shown in Fig.~\ref{fig:introduction}, human observers can assess the quality of the distorted image \(I_{d}\) with the guidance of an (or many) undistorted image(s) \(I_{er}\) even though their contents are not related in a noticeable way. In other words, it is intuitive to bridge the gap just by loosening the content restriction of reference images. 

In this paper, we aim to develop a new IQA scheme, \ie, external-reference IQA (ER-IQA). It utilizes external reference images to improve the performance of IQA algorithms in the no-reference scenario. Here, external reference images are defined as undistorted, high-quality images with arbitrary contents. We believe that natural images share common characteristics that can benefit the quality assessment procedure. Therefore, the proposed ER-IQA should achieve better performance than the NR ones while being just as practical. To the best of our knowledge, this is the first work focusing on quality assessment in the content-unpaired scenario. As the pioneer implementation, we introduce a simple baseline network to show the effectiveness of ER-IQA. Moreover, in order to utilize the reference information from unpaired input, a Mutual Attention-based Feature Enhancement (MAFE) module is proposed by exploring the relation between feature embeddings. 

Our main contributions can be summarized as follows:
\begin{itemize}
\item We propose ER-IQA, a novel IQA scheme that uses arbitrary images as references to enhance the performance of IQA models in the same no-reference scenario. To the best of our knowledge, this is the first work that explores the power of external reference images and the quality assessment in the unpaired scenario.
\item We design a quality assessment network for unpaired images (Unpaired-IQA) as the first implementation and a baseline of the ER-IQA scheme. Specifically, we propose a Mutual Attention-based Feature Enhancement (MAFE) module that allows the network to extract discriminative and robust features from distorted images and external reference ones, respectively.
\item Experimental results validate the effectiveness of ER-IQA and the superior performance of the proposed network. By exploiting external reference information, not only does our model outperform the state-of-the-art NR methods, but it also narrows the huge gap between FR and NR methods.
\end{itemize}

\section{Related Work}
\label{sec:rw}

\begin{figure*}
\begin{center}
   \includegraphics[width=0.95\textwidth]{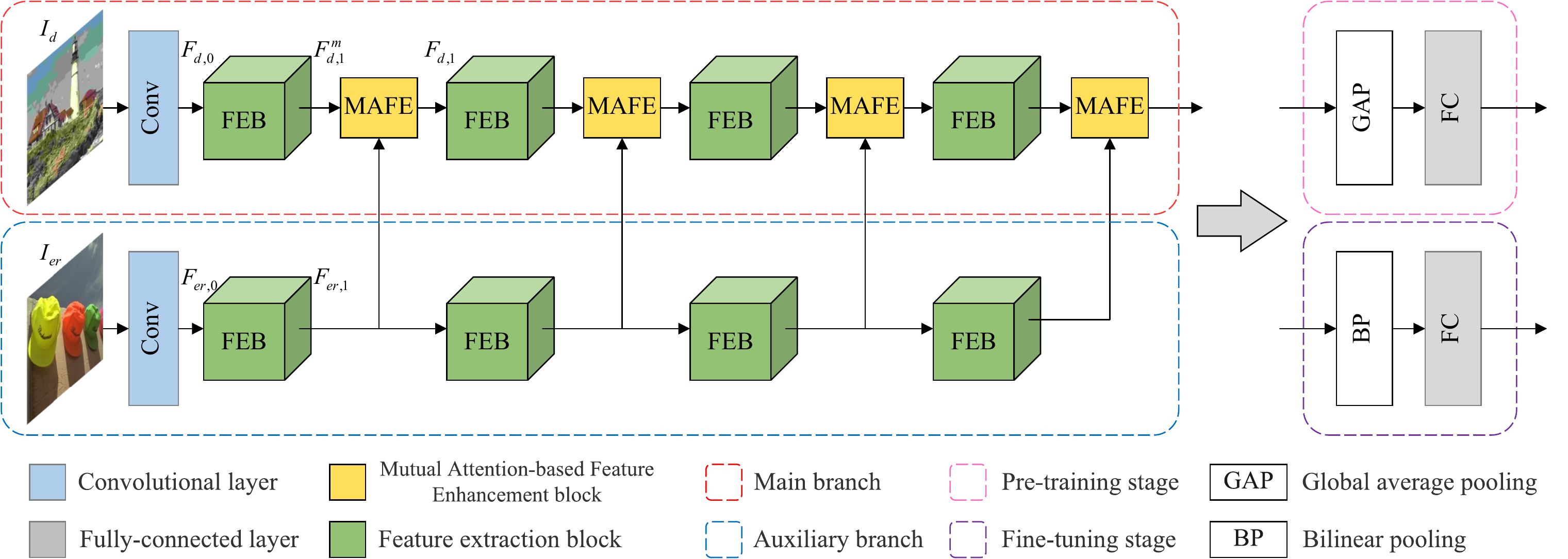}
\end{center}
   \caption{The main structure of our proposed Unpaired-IQA. The main branch and the auxiliary branch extract features from the distorted image and the external reference image, respectively, which are adaptively fused through the proposed MAFE module. Output of the last MAFE is used for classification during pre-training or sent into a bilinear pooling module to get final scores during fine-tuning.}
\label{fig:overview}
\end{figure*}
\subsection{No-reference Image Quality Assessment}

Conventional NR-IQA methods tend to extract hand-crafted features from images to be assessed. Natural scene statistics (NSS)-based models assume that natural, high-quality images have certain statistical characteristics sensitive to distortions and extract features in different domains, \eg, Wavelets \cite{moorthy2010two}, Discrete Cosine Transform (DCT) \cite{saad2012blind}, \etc. Though such approaches may be effective when facing certain known distortion types, they can hardly handle real-world scenes where multiple, complex distortions exist. 

With the great success of deep learning in vision tasks \cite{gong2017motion, yan2019multi}, CNNs are widely used in NR-IQA \cite{kang2014convolutional}, \cite{kang2015simultaneous}, \cite{bianco2018use}, \cite{wu2015blind}, \cite{kim2016fully}, \cite{ma2017end}. One key issue of CNN-based methods is the demand for large-scale labeled data, yet samples of existing IQA datasets are far from enough. To address this problem, Bianco \etal \cite{bianco2018use} fine-tuned a CNN pre-trained on ImageNet \cite{krizhevsky2012imagenet} for quality prediction. Talebi and Milanfar \cite{talebi2018nima} proposed a model also pre-trained on ImageNet and fine-tuned on IQA datasets by predicting the distribution of human opinion scores. In \cite{zhang2020blind}, Zhang \etal adopted the pre-trained VGG-16 model \cite{simonyan2014very} to handle authentic distortions and use it together with a CNN for synthetic distortions for final quality assessment. Moreover, Su \etal \cite{su2020blindly} sufficiently took advantage of image semantics extracted by a pre-trained ResNet \cite{he2016deep} and achieved significant success in handling authentic distortions. Other strategies, \eg, learning to rank \cite{liu2017rankiqa} and adopting generative adversarial networks (GANs) \cite{lin2018hallucinated}, are also proved effective for NR-IQA.

Although these methods have achieved promising performance, we argue the real challenge of NR-IQA lies in the insufficient supervision the labels (MOSs) can provide the learning process in the no-reference scenario. Unlike the FR scheme, where a quality score is defined as the difference or the similarity between a distorted image and its corresponding reference, scores in NR-IQA are less informative and meaningful. In this paper, we seek a novel scheme to tackle the above problem. Inspired by conventional methods, we also believe that natural, high-quality images share common characteristics, which can be extracted and utilized by CNNs to boost performance. Thus, we introduce external reference images for accurate quality prediction without loosening the pre-condition of NR-IQA. 

\subsection{Full-reference Image Quality Assessment}
Benefiting from the power of benchmark information, FR-IQA methods have achieved remarkable performance, which already leads to a significant gap between FR methods and NR ones concerning the consistency with human perception. FR-IQA methods have been widely applied as perceptual metrics to compute the similarity between images in a pixel level \cite{wang2004image}, \cite{sheikh2006image}, \cite{xue2013gradient}, or between feature embeddings in deep space \cite{zhang2018unreasonable}. However, the demand for the original, undistorted images limits FR-IQA's other applications \cite{zhang2020blind}, \cite{su2020blindly}. Note that there are some works that try to handle the situation when the reference image and the distorted one are not pixel-wise aligned \cite{10.1007/978-3-319-46454-1_1}. But they still require similar scenes as references, limiting their application in the no-reference scenario. 

In this paper, we aim to learn from the FR scheme when dealing with the more practical no-reference scenarios. We introduce a new IQA subcategory, namely external-reference IQA, to narrow the gap between the two existing IQA schemes.



\section{Our Approach}

In this section, we present our approach to the proposed ER-IQA. To demonstrate the potential of ER-IQA, we design a simple yet efficient network consisting of two branches, \ie, the main branch for distorted images and the auxiliary branch for external reference images. We also discuss several choices of fusion operation for the unpaired inputs, including the proposed MAFE module that is further proved the most effective. Furthermore, we discuss the relation between the proposed scheme and existing IQA categories to explain our motivation and the feasibility of the ER-IQA sub-category.

\subsection{Basic Network Architecture}

As we mentioned in the previous sections, the motivation of applying external reference for the distorted image is to compensate for the absence of the actual reference image in a no-reference scenario. Therefore, the proposed Unpaired-IQA network is designed to extract quality-related features from the distorted image \(I_{d}\) while adaptively collecting useful information from the unpaired, external reference image \(I_{er}\). As shown in Fig.~\ref{fig:overview}, our Unpaired-IQA network consists of two branches: the main branch for distorted images and an auxiliary branch for external reference images. The two branches have an identical structure yet do not share weights since they are expected to focus on different information inside \(I_{d}\) and \(I_{er}\). At the beginning of each branch, we use a $3\times3$ convolutional layer to extract features from the input image.
\begin{equation}\label{eq:1}
    \begin{aligned}
       F_{d,0}&=Conv_{3\times3}\left ( I_d \right ), \\
       F_{er,0}&=Conv_{3\times3}\left ( I_{er} \right ),
    \end{aligned}
\end{equation}
where \(Conv_{3\times3}\left ( \cdot \right )\) represents a $3\times3$ convolutional layer, and \(F_{d,0}\) and \(F_{er,0}\) serve as the inputs of next stage. 

Then deep features are extracted through the basic blocks that each consist of two convolutional layers with the Rectified Linear Unit (ReLU) \cite{glorot2011deep}.
\begin{equation} \label{eq:FEB}
	\begin{aligned}
		F_{d,i+1}^{m} & = FEB_{d,i}\left ( F_{d,i} \right ), \\
 		F_{er,i+1} & = FEB_{er,i}\left ( F_{er,i} \right ),
 	\end{aligned}
\end{equation}
where, taking the main branch for example, \(FEB_{d,i}\left ( \cdot \right )\) represents the feature extraction operation of the \(i\)-th block in the main branch, whose input feature is \(F_{d,i}\) and output is \(F_{d,i+1}^{m}\). Note that the superscript \(m\) indicates that \(F_{d,i+1}^{m}\) is an intermediate feature before the fusion operation and becoming \(F_{d,i+1}\), the input of the \(\left(i+1\right)\)-th basic block.

As a crucial component of our model, the fusion module is designed for the features of the unpaired images \(I_d\) and \(I_{er}\). It should adaptively extract and utilize valuable information to boost the representation ability of the feature embeddings. Specifically, the input of the \((i+1)\)-th FEB of the main branch is generated through Eq.~\ref{eq:MAFE}.  
\begin{equation} \label{eq:MAFE}
  F_{d,i+1} = MAFE\left (  F_{d,i+1}^{m}, F_{er,i+1} \right ),
\end{equation}
where \(MAFE\left( \cdot\right)\) denotes the function of our proposed MAFE module.

Outputs of the last fusion module are used for generating the final results. Similar to \cite{zhang2020blind}, we utilize knowledge transfer technique to handle images with authentic distortions. A network pre-trained for the image classification task on ImageNet \cite{krizhevsky2012imagenet} is adopted to extract authentic distortion related features and the output feature of its last convolutional layer. Subsequently, the two outputs are merged through bilinear pooling following \cite{zhang2020blind} before sent into the final regression layer.

\subsection{Adaptive Fusion for Unpaired Features} \label{Adaptive}

The challenge of ER-IQA mainly lies in the unpaired scenario of the distorted images and the reference ones concerning their contents. Although it is easy for human observers to assess the quality with external reference guidance, it remains challenging for algorithms. Hence, how to effectively extract and use the information inside features from external reference images is the key of ER-IQA. Here we introduce three different kinds of feature fusion designs, including our proposed MAFE.

\textbf{Cosine Similarity Guided Feature Fusion.} As we do not limit the contents of external reference images, which can be extremely various, a straightforward way to guide the feature fusion is to exploit the similarity between their corresponding embeddings. Here we choose cosine similarity as a hand-crafted similarity measure. For two features \(F^{m}_{d}\) from the main branch and \(F_{er}\) from the auxiliary branch, we calculate the cosine similarity between feature maps of the corresponding channels. Specifically, for feature maps from the \(i\)-th channel \(F_{d,m}^{\left (i \right)}\) and \(F_{er}^{\left (i \right)}\), the cosine similarity \(s_{i}\) is computed by: 
\begin{equation}\label{eq:cosine1}
s_{i} = \frac{F_{d,m}^{\left(i\right)} \cdot F_{er}^{\left(i\right)}}
{\left \| F_{d,m}^{\left(i\right)} \right \|_2 \cdot \left \| F_{er}^{\left(i\right)} \right \|_2 },
\end{equation}
where \(\left\|\cdot\right\|_2 \) denotes \(\ell_2\)-norm for a feature map. 

Then we treat the measure as weight and perform a weighted sum of \(F_{d,m}^{\left (i \right)}\)  and \(F_{er}^{\left (i \right)}\) as the fusion operation. 
\begin{equation}\label{eq:cosine2}
F_{d}^{\left(i\right)} = F_{d,m}^{\left(i\right)} + s_i\cdot F_{er}^{\left(i\right)},
\end{equation}

\textbf{Bottleneck Based Feature Fusion.} One other choice of feature fusion is to leave it to the network without any prior knowledge or hand-crafted operations. A bottleneck layer is adopted to fuse \(F_d^m\) and \(F_{er}\) into a new feature \(F_f\) while keeping the number of channels unchanged. 
\begin{equation}\label{eq:bottleneck}
F_d = Conv_{1\times 1} \left ( concat\left ( F_d^m, F_{er} \right )  \right ) ,
\end{equation}
where \(concat \left ( \cdot \right )\) denotes feature concatenation operation, \(Conv_{1\times 1}\left(\cdot\right)\) is a \(1\times 1\) convolution responsible for dimension reduction.

\begin{figure}[t]
\begin{center}
    \includegraphics[width=0.95\linewidth]{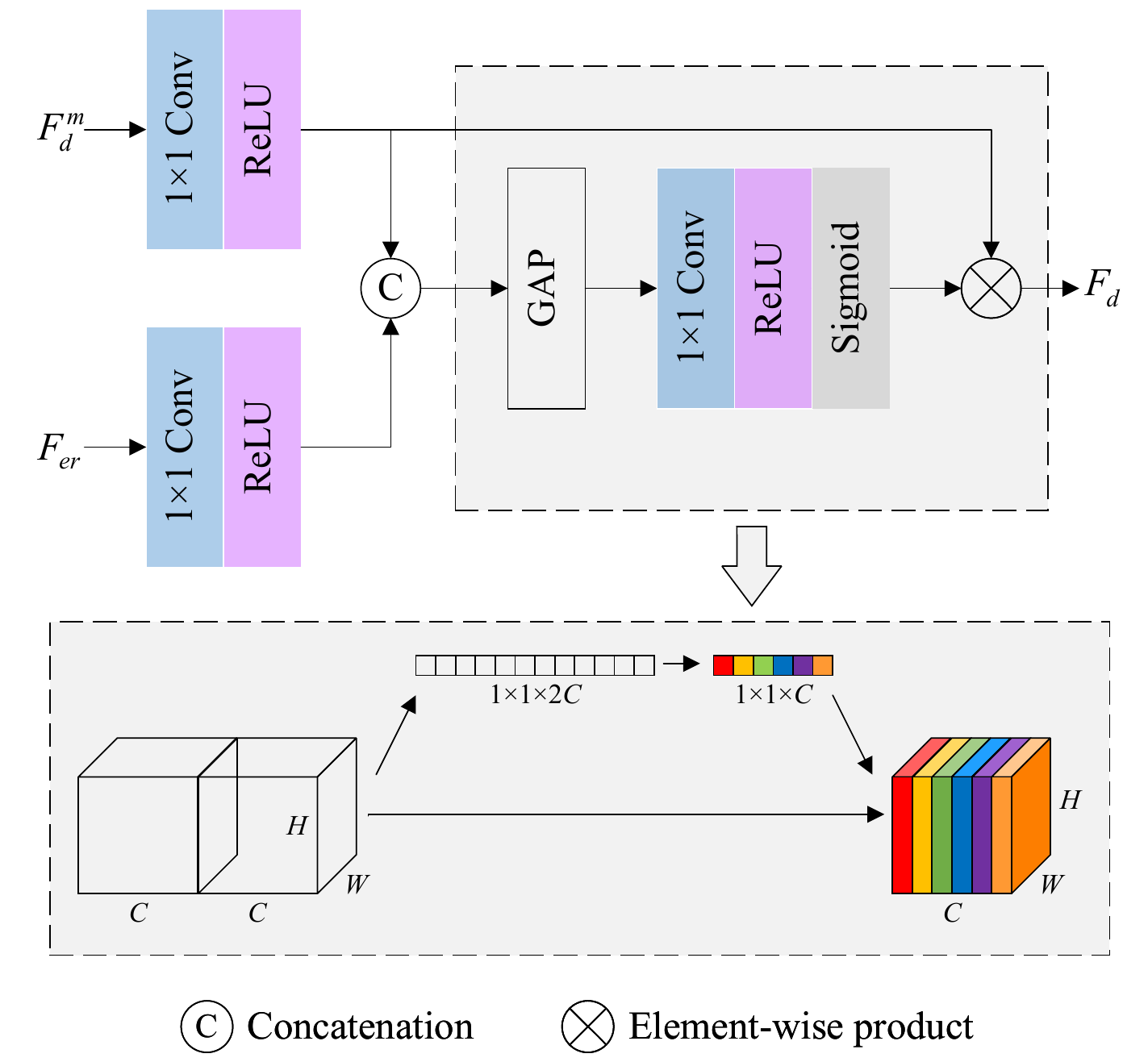}
\end{center}
   \caption{The proposed self-adaptive feature fusion module for unpaired features.}
\label{fig:MAFE}
\end{figure}

\textbf{Mutual Attention-based Feature Enhancement.} As mentioned before, the two branches of our Unpaired-IQA network share an identical structure. So given a pair of features, one straight idea is to extract useful information and abandon the redundant one selectively. We argue that utilizing existing similarity metrics, \eg, cosine similarity, to guide the process does not fully exploit the potential of deep learning. But simply placing a bottleneck layer without any interpretable design is just brute. Therefore, we propose a learnable module to realize self-adaptive feature enhancement for unpaired images.

Inspired by \cite{hu2020squeeze}, we promote a channel-wise feature enhancement guided by the reference feature extracted from external reference images through an attention-based manner. As depicted in Fig.~\ref{fig:MAFE}, the MAFE module takes two features as input and generates the enhanced feature without artificial operations. Two $1\times1$ convolutional layers, followed by ReLU \cite{glorot2011deep}, are first adopted to add some additional non-linearity before a channel-wise operation. Then the two resulting features are concatenated and pooled, and used to generate an attention vector, similar to the Squeeze-and-Excitation operation in \cite{hu2020squeeze}. The difference is that we only squeeze the intermediate vector with a factor of $2$, resulting in a halved channel number so that an element-wise product between the vector and $F_d^m$ can be applied. In this way, mutual attention is realized and can benefit the learning process.

The philosophy of our MAFE is that we not only model the channel inter-dependencies of the main feature $F_d^m$ but also focus on channel-wise inter-feature dependencies. We provide the feature embedding with access to the global information of both the distorted image and the external reference image, expecting the latter to further enhance the representation ability of the convolutional features.


\begin{table*}[tp]
\caption{Single dataset evaluation on five datasets. Weighted average is also presented.} \label{tab:single}
\begin{center}
\resizebox{\textwidth}{!}{%
\begin{tabular}{ccccccccccccc}
\toprule
\multirow{2}{*}{Method} &
  \multicolumn{2}{c}{LIVE \cite{sheikh2006statistical}} &
  \multicolumn{2}{c}{CSIQ \cite{larson2010most}} &
  \multicolumn{2}{c}{TID2013 \cite{ponomarenko2013color}} &
  \multicolumn{2}{c}{LIVEC \cite{ghadiyaram2015massive}} &
  \multicolumn{2}{c}{KonIQ \cite{hosu2020koniq}} &
  \multicolumn{2}{c}{WA} \\
                & SROCC          & PLCC           & SROCC & PLCC  & SROCC & PLCC  & SROCC & PLCC  & SROCC & PLCC  & SROCC & PLCC  \\ \midrule
BRISQUE \cite{mittal2012no}  & 0.939          & 0.942          & 0.750 & 0.829 & 0.573 & 0.651 & –     & –     & –     & –     & 0.667 &  0.733 \\
HOSA \cite{xu2016blind}     & 0.948          & 0.949          & 0.781 & 0.842 & 0.688 & 0.764 & –     & –     & –     & –     & 0.749 & 0.810 \\
BIECON \cite{kim2016fully}   & 0.958          & 0.960          & 0.815 & 0.823 & 0.717 & 0.762 & –     & –     & –     & –     & 0.776 & 0.807 \\
WaDIQaM \cite{bosse2017deep}  & 0.954          & 0.963          & –     & –     & 0.761 & 0.787 & –     & –     & –     & –     & 0.801 & 0.823 \\
BPSQM \cite{pan2018blind}    & 0.973          & 0.963          & 0.874 & 0.915 & 0.862 & 0.885 & –     & –     & –     & –     & 0.883 & 0.904 \\
DIQA \cite{kim2018deep}     & \textbf{0.975} & \textbf{0.977} & 0.884 & 0.915 & 0.825 & 0.850 & –     & –     & –     & –     & 0.861 & 0.883 \\
DB-CNN \cite{zhang2020blind}   & 0.968          & 0.971          & 0.946 & 0.959 & 0.816 & 0.865 & 0.851 & 0.869 & 0.875 & 0.884 & 0.871 & 0.888 \\
CaHDC \cite{wu2020end}    & 0.965          & 0.964          & 0.903 & 0.914 & 0.862 & 0.878 & 0.738 & 0.744 & –     & –     & 0.857 & 0.868 \\
HyperIQA \cite{su2020blindly}  & 0.962          & 0.966          & 0.923 & 0.942 & –     & –     & 0.859 & 0.882 & 0.906 & 0.917 & 0.906 & 0.918 \\ \midrule
Baseline & 0.959          & 0.967          & 0.939 & 0.945 & 0.812     & 0.850     & 0.857 & 0.877 & 0.870 & 0.881 & 0.866 & 0.883 \\ 
\textbf{Ours} &
  0.970 &
  0.973 &
  \textbf{0.948} &
  \textbf{0.960} &
  \textbf{0.865} &
  \textbf{0.885} &
  \textbf{0.864} &
  \textbf{0.886} &
  \textbf{0.929} &
  \textbf{0.941} &
  \textbf{0.915} &
  \textbf{0.929} \\ \bottomrule
\end{tabular}%
}
\end{center}
\end{table*}

\begin{table*}[]
\centering
\caption{Comparison with FR methods.}
\label{tab:single_FR}
\small
\begin{tabular}{ccccccccc}
\toprule
\multirow{2}{*}{Method} & \multicolumn{2}{c}{LIVE}  & \multicolumn{2}{c}{CSIQ}  & \multicolumn{2}{c}{TID2013} & \multicolumn{2}{c}{WA} \\
              & SROCC & PLCC  & SROCC & PLCC  & SROCC & PLCC           & SROCC & PLCC           \\ \midrule
SSIM \cite{wang2004image}         & 0.948 & 0.945 & 0.876 & 0.761 & 0.637 & 0.691          & 0.734 & 0.747          \\
MS-SSIM \cite{wang2003multiscale}      & 0.951 & 0.949 & 0.913 & 0.899 & 0.786 & 0.833          & 0.837 & 0.865          \\
VIF \cite{sheikh2006image}          & 0.963 & 0.960 & 0.920 & 0.928 & 0.677 & 0.772          & 0.770 & 0.833          \\
GMSD \cite{xue2013gradient}          & 0.960 & 0.960 & 0.957 & 0.954 & 0.804 & 0.859          & 0.859 & 0.894          \\
FSIMc \cite{zhang2011fsim}         & 0.962 & 0.962 & 0.932 & 0.920 & 0.851 & 0.877          & 0.885 & 0.899          \\
DeepQA \cite{kim2017deep_conf}       & 0.981 & 0.982 & 0.961 & 0.956 & 0.939 & \textbf{0.947} & 0.950 & \textbf{0.955} \\
WaDIQaM-FR \cite{bosse2017deep}   & 0.970 & 0.980 & -     & -     & 0.761 & 0.787          & 0.804 & 0.827          \\
DRF-IQA \cite{kim2020dynamic}                & \textbf{0.983} & \textbf{0.983} & \textbf{0.964} & \textbf{0.960} & \textbf{0.944}       & 0.942       & \textbf{0.954} & 0.952 \\ \midrule
Baseline      & 0.959 & 0.967 & 0.939 & 0.945 & 0.812 & 0.850          & 0.860 & 0.887          \\
\textbf{Ours} & 0.970 & 0.973 & 0.948 & 0.960 & 0.865 & 0.885          & 0.898 & 0.914          \\ \bottomrule
\end{tabular}%
\end{table*}

\begin{figure}[t]
\begin{center}
    \includegraphics[width=0.95\linewidth]{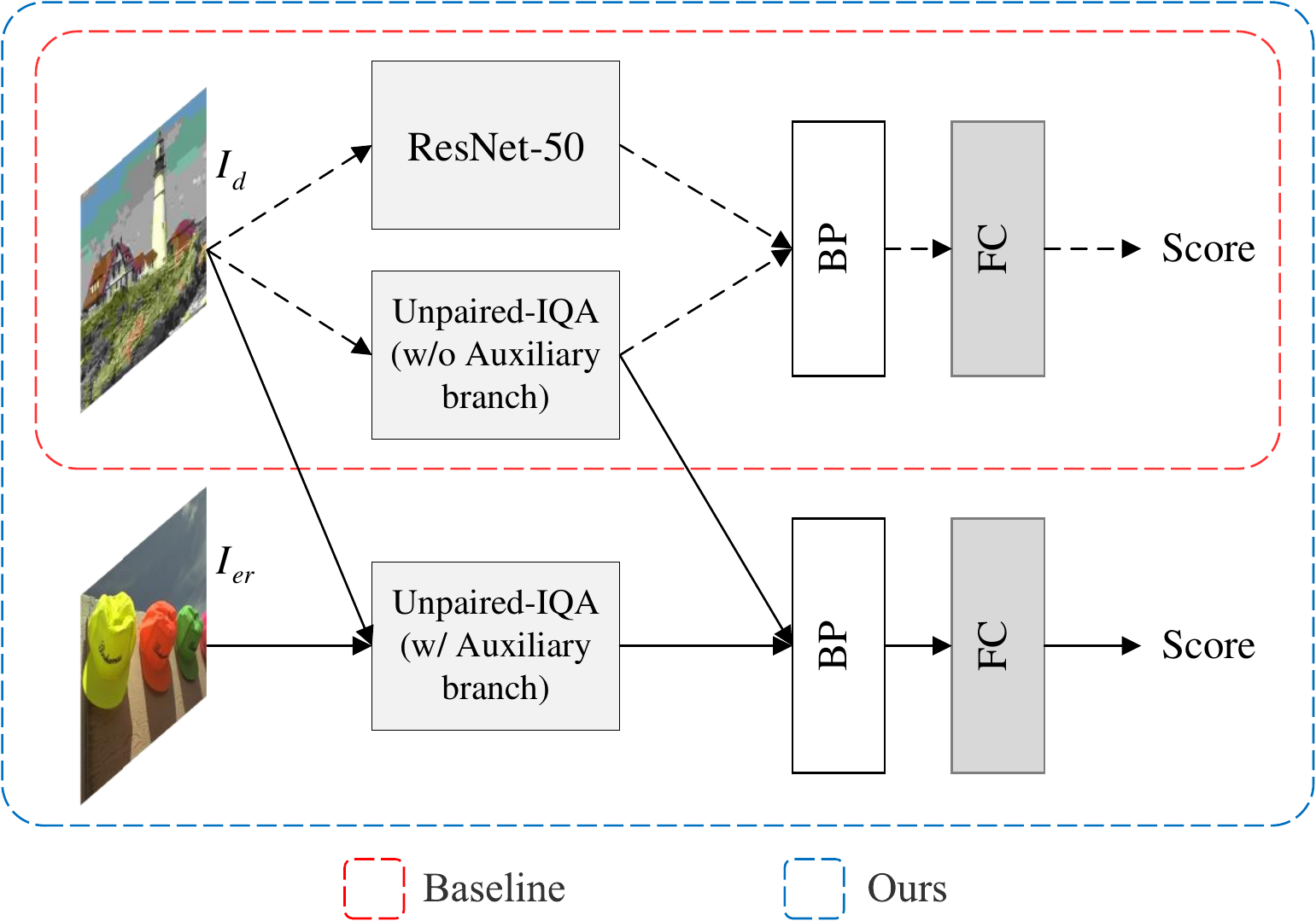}
\end{center}
    \caption{An illustration of the difference between our baseline and the proposed network.}
\label{fig:whole}
\end{figure}

\subsection{Discussions}
\label{Discussions}

\textbf{Difference to FR-IQA.} FR-IQA has always been the most consistent with the HVS among all existing categories. The key to its success is the essential benchmark, \ie, reference images, allowing FR models to achieve excellent performance only by learning to compare the differences between images. The main difference between FR-IQA and ER-IQA is the role of reference images. Reference images in ER-IQA are no longer considered benchmarks but additional inputs to boost assessment in the no-reference scenario. Contents of external reference images are not restricted, but on the contrary, arbitrary. Admittedly, this will for sure leads to lower performance. But we do not propose ER-IQA for the scenario where the original reference images are available. Instead, it is designed for the situation where it is difficult or even impossible to acquire the original, undistorted images so that FR methods are no longer applicable. After all, acquiring high-quality images with arbitrary contents is much easier. And we argue that these high-quality images share common characteristics that, when properly used, should benefit the quality assessment procedure. 

It is worth mentioning that the proposed Unpaired-IQA can also be applied to the FR-IQA task when offered content-paired inputs, yet may not achieve a promising performance compared to existing FR-IQA methods. This is because the underlying philosophy of the two schemes is fundamentally different. While the full-reference methods tend to utilize every little detail within reference images comprehensively, ER-IQA tries to alleviate the impact of content variation by selectively extract information. 

\textbf{Relation with NR-IQA.} ER-IQA is proposed under the same circumstance as NR-IQA, where the generally referred reference images are not available. It is motivated in the first place by existing NR-IQA methods \cite{yan2018two}, \cite{lin2018hallucinated} that introduce or generate additional prior information into the NR scheme for performance enhancement. Instead of using hand-crafted features as prior knowledge, We take one step forward and use externally acquired images as extra inputs for better performance. Users can choose any images they recognize as high-quality as external reference images, and the assessment is then expected to be more consistent with their own subjective opinions.

\subsection{Implementation Details}
\label{Implementation Details}

We choose the structure of DB-CNN \cite{zhang2020blind} as our baseline when evaluating our model's performance. As illustrated in Fig.~\ref{fig:whole}, the baseline model contains a ResNet-50 \cite{he2016deep} pre-trained on classification task and the only main branch of Unpaired-IQA without the auxiliary branch or MAFEs. This is basically an NR model following the design of DB-CNN only with different backbones. And then, we use the whole Unpaired-IQA to constitute our final model. It is worth mentioning that we only introduce the auxiliary branch into one of the sub-networks for simplicity. 

Following the procedure in \cite{zhang2020blind}, the Unpaired-IQA network is first pre-trained for a distortion classification task with unpaired inputs. Following \cite{zhang2020blind}, thirty-nine classes of distorted images are generated from pristine images chosen from Waterloo Exploration Database \cite{ma2016waterloo} and the DIV2K dataset \cite{agustsson2017ntire}. Here we add one more class for undistorted images, \ie, external reference images. Note that DB-CNN is originally trained on a complex dataset with manually chosen samples. To assure fair comparison, we retrain our baseline from scratch on our dataset and report results after fine-tuning for IQA. In our setting, for each distorted sample, its external reference image is randomly selected from all other pristine images to ensure the unpaired condition. All samples are \(224\times224\) pixel patches extracted from the original images and randomly flipped for augmentation. We used Adam \cite{kingma2014adam} optimizer to train our model for $30$ epochs, with a mini-batch of $64$. We set the initial learning rate \(1\times10^{-3}\), which is reduced by $10$ times every $10$ epochs. During pre-training, we minimize cross-entropy loss for the classification task. Specifically, given $N$ training samples in a mini-batch, loss is computed as: 
\begin{equation} \label{eq:cross}
\ell_{pre}= -\sum_{i=1}^{N} \sum_{j=1}^{40} p_{j}^{\left ( i \right ) } \log{\hat{p}_{j}^{\left ( i \right ) } },
\end{equation}
where \(p_{j}^{\left ( i \right )}\) is the ground-truth indicator of the \(i\)-th sample belonging to the \(j\)-th class, and \(\hat{p}_{j}^{\left ( i \right )}\) is the predicted probability of the \(i\)-th input.

During fine-tuning on the IQA task, the size of a mini-batch is set to $8$. We train our model for $100$ epochs with the learning rate set to \(1\times10^{-5}\). The input size of image samples is adjusted depending on the specific dataset. We minimize the mean squared error (MSE) loss over the training set:
\begin{equation} \label{eq:l2}
\ell= \frac{1}{N} \sum_{i=1}^{N} \left ( y_{i} - \hat{y}_{i} \right ) ^{2},
\end{equation}
where \(N\) is the number of samples in a mini-batch, \(y_{i}\) denotes the ground-truth score of the i-th sample, and \(\hat{y}_{i}\) is the predicted score.

Our model is implemented based on PyTorch, and all experiments are performed on NVIDIA 1080Ti GPUs. 

\section{Experiments}

\subsection{Datasets}
We perform experiments on several IQA benchmark datasets. Three synthetic image datasets, \ie, LIVE \cite{sheikh2006statistical}, CSIQ \cite{larson2010most}, and TID2013 \cite{ponomarenko2013color}, together with two authentic image datasets, \ie, LIVE Challenge (LIVEC) \cite{ghadiyaram2015massive} and KonIQ-10k \cite{hosu2020koniq}, are used for experiments. LIVE \cite{sheikh2006statistical} contains $779$ distorted images synthesized from $29$ reference images covering five distortion types, \ie, Gaussian blur (GB), Gaussian noise (WN), JPEG compression (JPEG), JPEG2000 compression (JP2K), and fast fading (FF). Similarly, CSIQ \cite{larson2010most} is composed of $866$ distorted images with six types of distortions, \ie, JPEG, JP2K, GB, WN, contrast change (CG), and pink noise (PN) generated from $30$ reference images. TID2013 \cite{ponomarenko2013color} dataset consists of $25$ reference images and $3000$ distorted images with twenty-four distortion types at five degradation levels. LIVEC \cite{ghadiyaram2015massive} contains $1162$ images obtained from the real world containing widely diverse authentic distortions. KonIQ-10k \cite{hosu2020koniq} consists of $10073$ images selected from ten million entries diverse in content and distortions. 

\begin{table}[tp]
\caption{SROCC results of individual distortion types on the LIVE dataset.} \label{tab:type}
\begin{center}
\small
\begin{tabular}{cccccc}
\toprule
Method   & JPEG           & JP2K           & WN             & GB             & FF             \\ \midrule
BRISQUE \cite{mittal2012no}  & 0.965          & 0.929          & 0.982          & 0.964          & 0.828          \\
HOSA \cite{xu2016blind}    & 0.954          & 0.935          & 0.975          & 0.954          & 0.954          \\
BIECON \cite{kim2016fully}  & \textbf{0.974} & 0.952          & 0.980          & 0.956          & 0.923          \\
WaDIQaM \cite{bosse2017deep} & 0.953          & 0.942          & 0.982          & 0.938          & 0.923          \\
BPSQM \cite{pan2018blind}   & 0.929          & \textbf{0.972} & \textbf{0.985} & \textbf{0.977} & \textbf{0.964} \\
DB-CNN \cite{zhang2020blind}  & 0.972          & 0.955          & 0.980          & 0.935          & 0.930          \\
HyperIQA \cite{su2020blindly} & 0.961          & 0.949          & 0.982          & 0.926          & 0.934          \\ \midrule
Ours     & \textbf{0.974} & 0.952          & \textbf{0.985} & 0.940          & 0.939          \\ \bottomrule
\end{tabular}%
\end{center}
\end{table}

\begin{table}[tp]
\begin{center}
\caption{SROCC results of cross dataset tests.} \label{tab:cross}
\resizebox{\linewidth}{!}{%
\small
\begin{tabular}{cccccc}
\toprule
 & Training & Testing & DB-CNN & HyperIQA & Ours \\ \midrule
\multirow{2}{*}{Authentic} & LIVEC & KonIQ & 0.754 & 0.772 & \textbf{0.806} \\
                          & KonIQ & LIVEC & 0.755 & 0.785 & \textbf{0.809} \\ \midrule
\multirow{2}{*}{Synthetic} & LIVE  & CSIQ  & 0.758 & 0.744 & \textbf{0.871} \\
                          & CSIQ  & LIVE  & 0.877 & 0.926 & \textbf{0.928} \\ \midrule
\multirow{2}{*}{A. \& S.}    & LIVEC & LIVE  & 0.746 & –     & \textbf{0.785} \\
                          & LIVE  & LIVEC & 0.567 & –     & \textbf{0.591} \\ \bottomrule
\end{tabular}
}%
\end{center}
\end{table}
\begin{table}[tp]
\caption{D-Test, L-Test and P-Test results on the Waterloo Exploration Database.} \label{tab:dtp}
\begin{center}
\small
\resizebox{0.7\linewidth}{!}{
\begin{tabular}{cccc}
\toprule
Method           & D-Test          & L-Test          & P-Test          \\ \midrule
BRISQUE \cite{mittal2012no}    & 0.9204          & 0.9772          & 0.9930          \\
CORNIA \cite{ye2012unsupervised}     & 0.9290          & 0.9764          & 0.9947          \\
HOSA \cite{xu2016blind}       & 0.9175          & 0.9647          & 0.9947          \\
WaDIQaM \cite{bosse2017deep}     & 0.9074          & 0.9467          & 0.9628          \\
dipIQ \cite{ma2017dipiq}      & 0.9346          & \textbf{0.9846} & \textbf{0.9999} \\
MEON \cite{ma2017end}       & 0.9384          & 0.9669          & 0.9984          \\
DB-CNN \cite{zhang2020blind}   & \textbf{0.9402} & 0.9448          & 0.9980          \\
HyperIQA \cite{su2020blindly} & 0.9006          & 0.9747          & 0.9971          \\ \midrule
Ours            & 0.9307          & 0.9541          & 0.9979          \\ \bottomrule
\end{tabular}
}%
\end{center}
\end{table}
\subsection{Experiment Protocols and Criteria}

Spearman's rank order correlation coefficient (SROCC) and Pearson's linear correlation coefficient (PLCC) are employed as evaluation criteria. Given \(N\) distorted images, the SROCC is computed as:
\begin{equation}\label{eq:SROCC}
SROCC= 1- \frac{6 {\textstyle \sum_{i= 1}^{N}d_{i}^{2}  } }{N\left ( N^{2} - 1 \right ) },
\end{equation}
where \(d_{i}\) denotes the difference between the ranks of \(i\)-th test image in ground-truth and predicted quality scores. Before computing PLCC, we adopt the logistic function \cite{zhang2011fsim} for nonlinear regression as follows:
\begin{equation} \label{eq:non-lin}
    p = \beta_1\left ( \frac{1}{2} - \frac{1}{1+e^{\beta_2\left (q - \beta_3  \right )} }  \right ) + 
\beta_4q + \beta_5
\end{equation}
where $q$ represents the results of an IQA method, $p$ denotes the regression values of $q$, and $\beta_i \left ( i = 1, 2, 3, 4, 5\right )$ are parameters to be fitted. After the regression, the PLCC is computed as:
\begin{equation} \label{eq:PLCC}
PLCC= \frac{ {\textstyle \sum_{i=1}^{N}} \left ( y_{i}- \mu_{y}   \right )\left ( \hat{y} _{i}- \mu_{\hat{y}} \right )  }
{\sqrt{ {\textstyle \sum_{i=1}^{N}} \left ( y_{i}- \mu_{y}  \right )^{2} } \sqrt{ {\textstyle \sum_{i=1}^{N}} \left ( \hat{y} _{i}- \mu_{\hat{y}} \right )^{2} } },
\end{equation}
where \(\mu_{y}\) and \(\mu_{\hat{y}}\) are the means of the ground truth and predicted quality scores, respectively, \(y_{i}\) denote the ground-truth score of \(i\)-th image, and the predicted score from the network is \(\hat{y}_{i}\).

We conduct experiments by following the same protocol in \cite{zhang2020blind}. For each dataset, $80\%$ of images are used for training, and the rest $20\%$ are for testing. For synthetic datasets LIVE and CSIQ, the split is implemented according to reference images to avoid content overlapping. And for each distorted sample, the external reference is randomly selected from all the other reference images within the split set, other than its corresponding one, to meet the unpaired setting and prevent content overlapping of external references during training and testing. We run $10$ times of this random train-test splitting operation, and the median result are reported. 
\subsection{Comparison with the State-of-the-arts}

\begin{figure*}[!tp]
\begin{center}
    \includegraphics[width=0.95\textwidth]{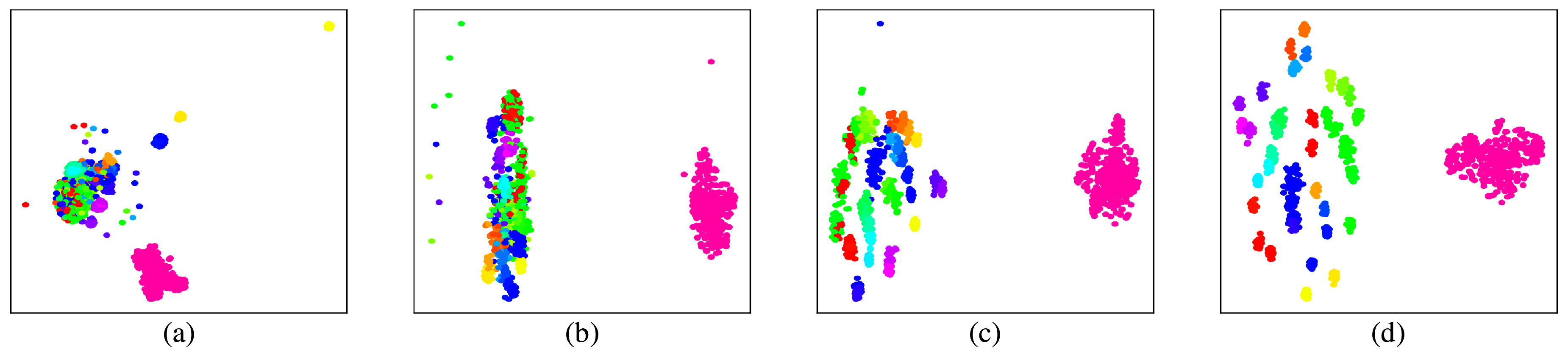}
\end{center}
   \caption{Visualization of the projected intermediate features before each MAFE module (from (a) the first MAFE  to (d) the last one) of the Unpaired-IQA network, indicating discriminative and robust feature extraction. Best viewed in color. Note that the \textbf{\textcolor[RGB]{255,9,159}{pink}} points represent the features of external reference images.}
\label{fig:visual}
\end{figure*}

\textbf{Single dataset evaluations.} We first compare our ER-IQA model against the state-of-the-art NR-IQA methods: BRISQUE \cite{mittal2012no}, HOSA \cite{xu2016blind}, BIECON \cite{kim2016fully}, WaDIQaM \cite{bosse2017deep}, BPSQM \cite{pan2018blind}, DIQA \cite{kim2018deep}, DB-CNN \cite{zhang2020blind}, CaHDC \cite{wu2020end} and HyperIQA \cite{su2020blindly}, on single datasets, as shown in Table~\ref{tab:single}. The best method for each dataset is indicated in bold. For each method, the weighted average (WA) of its results is also shown in the last column for concise. The weight of each dataset is equal to the number of distorted samples in it. Here we can see that our model outperforms all the state-of-the-art NR methods concerning overall results. And it outperforms the baseline in all datasets. Specifically, our model performs the best on authentic image datasets and achieves comparable results on synthetic ones benefiting from external references. This proves the effectiveness of our approach and the proposed ER-IQA. We further evaluate the performance of our model on individual distortion types. As shown in Table~\ref{tab:type}, benefiting from naturalness information provided by external reference images, our method outperforms the original baseline method DB-CNN \cite{zhang2020blind} and achieves competing performances on individual distortion types.

We also compare the ER-IQA model against FR-IQA methods: SSIM \cite{wang2004image}, MS-SSIM \cite{wang2003multiscale}, VIF \cite{sheikh2006image}, GMSD \cite{xue2013gradient}, FSIMc \cite{zhang2011fsim}, DeepQA \cite{kim2017deep_conf}, WaDIQaM-FR \cite{bosse2017deep} and DRF-IQA \cite{kim2020dynamic}. As shown in Table~\ref{tab:single_FR}, the ER-IQA can achieve comparable results and thus narrow the gap between NR and FR-IQA, though it cannot outperform the state-of-the-art ones. But as mentioned in Section~\ref{Discussions}, ER-IQA is not proposed with the purpose of being applied in full-reference scenarios.

\textbf{Cross dataset evaluations.} Cross dataset evaluations are typically used for robustness tests, where a robust IQA method is expected to perform well not just on the training dataset but also on other IQA datasets. Here we evaluate the generalizability of our proposed ER-IQA through cross evaluations. We choose the two most competing approaches, \ie, DB-CNN \cite{zhang2020blind} and HyperIQA \cite{su2020blindly}, for comparison. As there are no reference images in authentic image datasets, we use the ones provided by a synthetic image dataset, \ie, TID2013 \cite{ponomarenko2013color}, as external references. As shown in Table~\ref{tab:cross}, our model outperforms the other methods among all six cross evaluations, showing significant robustness when handling both real distortions and synthetic ones.

The cross dataset evaluations can also demonstrate the feasibility of the arbitrary choice of external reference images. For example, results on authentic datasets, \ie, LIVEC \cite{ghadiyaram2015massive} and KonIQ-10k \cite{hosu2020koniq}, reveal that when the external reference images are widely accepted as undistorted, high-quality ones, our model can extract and benefit from the useful information adaptively despite the varying, arbitrary content. When conducting experiments on synthetic datasets, we ensure no content overlapping of reference images among the training and testing sets. The reference images are chosen within the same dataset, yet samples are still in an unpaired manner through multiple shuffle and augmentation strategies. In this case, the experiments simulate the situation when users (testing samples) have slightly different subjective opinions from the ones carried in training samples, practically evaluate the controllability and generalization ability. As presented in Table~\ref{tab:cross}, the proposed ER-IQA model can still overcome the gap between the training labels and the testing ones and thus significantly outperforms the two most state-of-the-art NR methods. As a result, the power of external reference images and the excellent controllability of the Unpaired-IQA network are verified.

\textbf{Generalization performance evaluations.} To further examine the generalization ability of the Unpaired-IQA model, we use three criteria, \ie, Pristine/Distorted Image Discriminability Test (D-Test), List-wise Ranking Consistency Test (L-Test), and Pairwise Preference Consistency Test (P-Test). For our Unpaired-IQA and DB-CNN \cite{zhang2020blind} that require pre-training, we first retrain them on samples generated from DIV2K \cite{agustsson2017ntire} datasets only to ensure content independence during training and testing. The models are then fine-tuned on LIVE \cite{sheikh2006statistical} dataset for the IQA task and tested on the Waterloo Exploration Database \cite{ma2016waterloo}. As shown in Table~\ref{tab:dtp}, despite the varying contents of external reference images, our approach still achieves competing performance.

\begin{table}[tp]
\caption{SROCC results of ablation experiments on LIVE and KonIQ datasets as investigations of the proposed MAFE and the ER-IQA scheme.} \label{tab:Ablation}
\begin{center}
\small
\begin{tabular}{|c|c|c|c|}
\hline
Auxiliary branch          & Fusion type    & LIVE           & KonIQ          \\ 
\hline    \hline
$\times$               & $\times$   & 0.959          & 0.870          \\ 
\hline
\multirow{3}{*}{$\surd$} & Cosine     & 0.966          & 0.911          \\ 
\cline{2-4} 
                       & Bottleneck & 0.961          & 0.915          \\ 
                       \cline{2-4} 
                       & MAFE       & \textbf{0.970} & \textbf{0.929} \\ 
                       \hline
\end{tabular}%
\end{center}
\end{table}
\subsection{Ablation study}

To investigate the efficiency of the proposed MAFE module, we conduct ablation experiments on the LIVE and KonIQ-10k datasets. The results are shown in Table~\ref{tab:Ablation}. 

We first evaluate the effectiveness of the proposed ER-IQA scheme. As mentioned in Section~\ref{Implementation Details}, we remove the auxiliary branch and the MAFE modules in Fig.~\ref{fig:overview} and use the resulting model as our baseline. One can observe remarkable improvements in the SROCC results on both datasets when external reference images are introduced. With the proposed MAFE module, our Unpaired-IQA obtains $1.1\%$ and $6.8\%$ improvements on the two datasets, respectively, compared to the baseline model. 

We then examine different fusion techniques introduced in \ref{Adaptive}. As the performance of feature fusion guided by cosine similarity and the one based on bottleneck competes with each other, the proposed MAFE module achieves the best with at least $0.4\%$ increase on the LIVE dataset and $1.5\%$ on KonIQ-10k, showing the promising performance of our design.

\subsection{Visualization}

To demonstrate Unpaired-IQA's ability of discriminative feature extraction qualitatively, we perform visualization of intermediate features before each MAFE module using t-SNE \cite{van2008visualizing} after pre-training. It is worth mentioning that all the external reference samples presented here contain different contents. As shown in Fig.~\ref{fig:visual}, as features of different distortion types are discriminated against each other, our model can also extract robust features from external reference images despite the varying contents with the proposed MAFE. This shows that the structure is capable of handle the distortion type classification task even when fed with unpaired inputs.

\section{Conclusion}

In this paper, we propose a new IQA scheme, \ie, external-reference image quality assessment, to boost the performance in the no-reference scenario. We design ER-IQA as a pioneer work of the newly proposed category. With the proposed MAFE module, Unpaired-IQA can adaptively perceive useful information from the user-supplied, content-arbitrary external reference image(s) and benefit the quality assessment. Extensive experiments demonstrate the superior performance of the Unpaired-IQA model and the ER-IQA scheme, thus opening up new prospects of practical applications of IQA. In the future, we plan to explore further the power of external reference images and how the choice of different reference images will affect the performance of IQA.



\ifCLASSOPTIONcaptionsoff
  \newpage
\fi



\bibliographystyle{./IEEEtran}
\bibliography{./IEEEabrv,./mybib}
%

%








\end{document}